\documentclass[letterpaper, 10 pt, conference]{IEEEtran}  

\IEEEoverridecommandlockouts                              
\usepackage{graphicx} 
\usepackage{amssymb}  
\usepackage{theorem}  
\usepackage{psfrag}  
\usepackage{bm}  
\usepackage{algorithmic}
\usepackage{cite}
\usepackage{amsmath}
\usepackage{algorithm}
\usepackage{array}
\usepackage{mdwmath}
\usepackage{mdwtab}
\usepackage{eqparbox}
\usepackage{bm}   
\usepackage{subfigure}
\usepackage[dvips]{color}

\newcommand{\qed}{\nobreak \ifvmode \relax \else
      \ifdim\lastskip<1.5em \hskip-\lastskip
      \hskip1.5em plus0em minus0.5em \fi \nobreak    
      \vrule height0.5em width0.5em depth0.25em\fi}  
      
      \newcommand{\ddt}{\frac{\textnormal{d}}{\textnormal{dt}}}

\title{\LARGE  Classical Model Predictive Control\\ of a Permanent Magnet Synchronous Motor }
\author{Jean-Fran\c cois Stumper, Alexander D\"otlinger and Ralph Kennel
}

\begin{document}

\maketitle

\section*{Keywords}

Optimal Control, Control of Drive, Synchronous Motor.

\section*{Address}

Institute of Electrical Drive Systems and Power Electronics, Technische Universit\"at M\"unchen, Germany.

\section*{Abstract}

A model predictive control (MPC) scheme for a permanent-magnet synchronous motor (PMSM) is presented. The torque controller optimizes a quadratic cost consisting of control error and machine losses repeatedly, accounting the voltage and current limitations. The scheme extensively relies on optimization, to meet the runtime limitation, a suboptimal algorithm based on differential flatness, continuous parameterization and linear programming is introduced. 

The multivariable controller exploits cross-coupling effects in the long-range constrained predictive control strategy. The optimization results in fast and smooth torque dynamics while inherently using field-weakening to improve the power efficiency and the current dynamics in high speed operation. As distinctive MPC feature, constraint handling is improved, instead of just saturating the control input, field weakening is applied dynamically to bypass the voltage limitation. The performance of the scheme is demonstrated by experimental and numerical results.

\section{Introduction}

The efforts of implementing predictive controllers in electrical drives aim at replacing the classical cascaded field-oriented control structure with PI controllers. The machine can be better exploited by improved control behavior, the system variables are optimized. In this contribution, the conventional torque and current control structure of two separate controllers is changed to multi-input multi-output (MIMO) control. By transformation into the field-oriented frame, torque generation is decoupled from flux variation, however, the current dynamics are still strongly coupled, therefore a MIMO controller is advantageous. The aimed improvements are better current and voltage constraint handling by exploiting cross-coupling between the orthogonal components, and better power efficiency and dynamics by optimally adjusting the currents in both dynamic and steady-state operation.

The major obstacle in implementing predictive control schemes is the limited computational power, inherited by the high sampling rates. The most widespread schemes trade computational feasibility against compromises in the problem formulation, for instance, generalized predictive control (GPC) has a high prediction horizon but is unconstrained, whereas predictive torque control (PTC) is constrained but so far only reaches $2$ steps of prediction \cite{Cortes}. To obtain the advantages claimed by classical MPC on constrained MIMO systems, both, inclusion of constraints and a high prediction horizon are required. Only few schemes so far satisfy these two requirements \cite{Geyer}. Using continuous control in the field-oriented frame, the analytical problem description enables using efficient optimization algorithms to maximize the obtained information for a given computational power. 

The online solution of the linearly constrained linear-quadratic problem, typical for MPC, requires quadratic programming (QP) algorithms, which are, however, computationally too expensive for drive systems. A recent development is the use of explicit MPC, where an offline solution is computed and stored as look-up table in the real-time controller \cite{Kuehl,inhFW}. The scheme reaches $5$ prediction steps with constraints. Recently an online algorithm based on a fast-gradient algorithm was proposed for reference tracking control of a grid-connected inverter \cite{onlineMPC}. While the runtime results are convincing, experiments are pending. The advantage of online optimization in MPC is the possibility to manipulate or adapt parameters, which results in simplified commissioning, an aspect that is becoming more and more important in the recent developments \cite{Lee}.

This implementation entitled 'classical MPC' relies extensively on online-optimization. It is embedded in a flatness-based predictive control scheme \cite{Fliess,Hag}, where trajectory optimization and prediction are decoupled, and the trajectory generation can be simplified \cite{Guay}. A suboptimal trajectory generation algorithm \cite{SK10} is applied which is based on a continuous approach, the variables are not discretized but represented as a polynomial with undetermined coefficients. The cost function is linearized in the unconstrained optimum in order to use a linear programming (LP) solver, which is amongst the simplest and fastest numerical optimizers. As result, a (suboptimal) prediction of $2$ ms with current and voltage constraints is obtained at $8$ kHz sampling rate. This way, by solving the computational problem without compromising the MPC formulation, the merits of this controller can be studied.

The initial results have been presented in \cite{EPEpaper}.


\section{Problem Statement}


\subsection{Machine Model}

In a fist step the machine model is linearized. This will result in a linear-quadratic optimization problem with linear constraints, a very common problem in optimal control systems \cite{Pierre} that is also simpler to solve in real-time. Assuming that the rotor speed does not change too much over the optimization horizon $T$,
\begin{align}
\frac{d}{dt} \omega_M(t) &\approx 0  \;\; \Rightarrow \omega_M(t) = \textnormal{const. } \;\; \forall t\in[0,T] ,
\end{align}
the PMSM model and the voltage equations become linear. The assumption is justified if the current control loop is faster than the speed control loop. The electrical subsystem of the machine, consisting of the quadrature and direct currents $i_q$ and $i_d$ (peak values), is given as
\begin{align}
L_d \ddt i_d &=  -R i_d  + n_p \omega_M L_q i_q  + u_d   ,                 \label{eq:id} \\
L_q \ddt i_q &=  -R i_q  - n_p \omega_M L_d i_d  -n_p \omega_M K  + u_q  , \label{eq:iq} \\
      \tau_M &=   \frac{3}{2} n_p K i_q .
\end{align}
The nomenclature is shown in table II in the appendix. It is noted that the reluctance torque $\tau_M^R = \frac{3}{2} n_p (L_d-L_q)i_di_q$ is neglected, as this term is very small compared to the electromagnetic torque in surface-mounted PMSMs or in machines with small saliency. Furthermore, it would render the model nonlinear, requiring nonlinear optimization methods \cite{Delaleau}.

\subsection{Optimization Goals and Cost Functional}

The formulation of a suitable cost functional is a key point in predictive control, as it is the only tuning possibility of the control scheme. The optimization is aiming at minimizing the control error for good dynamical performance as well as machine losses for better efficiency. Both goals are included in the cost functional. By choosing the cost functional and weights well, it is possible to find a good trade-off between both goals during transients, and eventually to fulfill both goals in steady-state. The cost functional for the predictive torque controller is
\begin{align}
 J = \int_0^T \left( P_{ctrl}(t) + W_L \cdot P_{loss}(t) \right) dt + T \cdot P_{ctrl}(T) \label{eq:J}
\end{align}
which trades off the squared control error from the constant torque reference $\tau_M^*$
\begin{align}
P_{ctrl}(t) = (\tau_M-\tau_M^*)^2
\end{align}
with machine losses 
\begin{align}
P_{loss}(t) &= \frac{3}{2}R(i_d^2+i_q^2) +\frac{3}{2} n_p\omega_M k_{Fe} (\Psi_d^2+\Psi_q^2) \notag\\
            = &\frac{3}{2}R(i_d^2+i_q^2) +\frac{3}{2} n_p\omega_M k_{Fe} ((L_d i_d+K)^2+ (L_q i_q)^2)   .
\end{align}
The first term in $P_{loss}$ represents copper losses, and the second term represents the iron losses consisting of hysteresis losses. The last term in (\ref{eq:J}) is the end-weight of the control error, it serves to improve convergence within finite time, especially when constraints are active. Its effect is shown in the results section. Eddy current losses are negligible on the tested machines, however, they could be included using the model presented in \cite{Eisenmodell}. The machine losses can be reduced by field-weakening, where a trade-off between copper and iron losses is found \cite{PMSMeff}. Imposing a negative direct current $i_d$, the flux magnitude in the stator is reduced while the copper losses increase. As the iron loss constant $k_{Fe}$ is not part of standard motor parameters, it has to be determined experimentally \cite{Eisenverlust}. The iron losses are quite considerable, for instance, the applied $2.64$ kW machine has about $43$ W copper losses and $200$ W iron losses in rated operation, therefore the parameter $k_{Fe}$ can be determined using only standard equipment. Inverter losses remain undiscussed as the switching frequency is arbitrary fixed.  

\begin{figure*}[!ht]
  \centering
  \includegraphics[width=13cm]{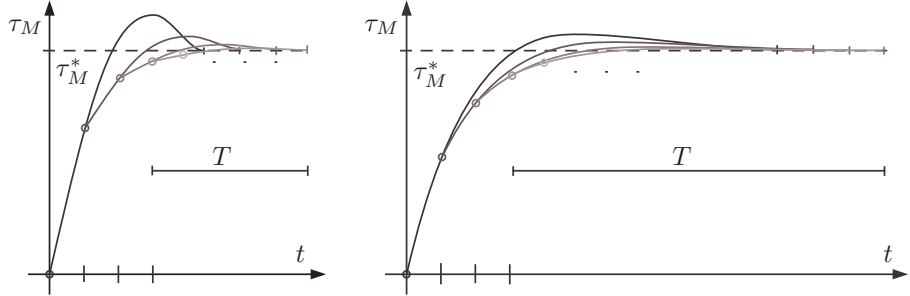}
  \caption{Exemplary torque setpoint change to describe open- and closed-loop trajectories in long-range predictive control. Left: small horizon, right: high horizon. Circles: starting points for each trajectory planning iteration. \label{fig:horizon}}
\end{figure*}

There are two tuning parameters, $W_L$ and $T$. The weight $W_L$ was set $0.05$, the value was determined heuristically. The optimization horizon is set $T=2$ ms such that the cost functional includes the complete setpoint change. It is important that the optimization horizon is high enough, otherwise the open-loop and closed-loop trajectories differ and the behavior is strongly suboptimal. This is illustrated in Fig. \ref{fig:horizon} (left). The open-loop trajectory, a continuous trajectory optimal with respect to $J$, is generated and a first part is applied to the system. At the next sampling step (indicated by a circle), a new open-loop trajectory is generated and again its first part is applied to the system. The resulting closed-loop trajectory, which is the sequence of the discrete points at each sampling interval, differs significantly from the open-loop trajectories, especially if the horizon is too small. Then, the closed-loop trajectories simply do not fit the cost functional anymore and are suboptimal. For a horizon higher than required for the setpoint change, as shown in Fig. \ref{fig:horizon} (right), the difference between open- and closed-loop trajectories becomes smaller. Ideally the points should be superposed with the continuous open-loop trajectories, then, the closed-loop trajectories can be assumed optimal regarding the open-loop cost functional $J$.

\psfrag{#y}{$\tau_M$}
\psfrag{#y*}{$\tau_M^*$}
\psfrag{#t}{$t$}
\psfrag{#T}{$T$}

\subsection{Current and Voltage Constraints}

The most important nonlinearities of a PMSM, in terms of control, are the  voltage and current limitations. The current constraints prevent overheating of the machine, and the voltage is limited by the maximum output voltage of the voltage source inverter. The voltage constraints limit rotor speed as well as current dynamics in high-speed operation. Both constraints are approximated by affine equations, in order to be computationally efficiently treated. 

\begin{figure}[!ht]
  \centering
  \includegraphics[width=8.5cm]{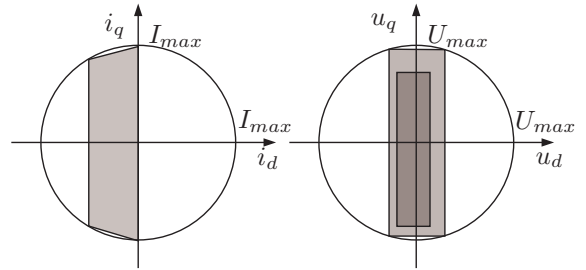}
  \caption{Affine approximation of the current and voltage constraints. Circle: feasible set of current and voltage vectors, grey: feasible set after approximation of the constraints.\label{fig:constraints}}
\end{figure}

The current range for the direct current $i_d$ is limited to $i_d^{min}\le i_d \le 0$. Only negative values of $i_d$ are desirable, as they improve power efficiency and reduce the induced voltage by weakening the flux magnitude in the stator \cite{PMSMeff,inhFW}. The lowest value $i_d^{min}$ is the optimum value at rated speed ($\frac{\partial}{\partial i_d} P_{loss}=0$) and is given as 
\begin{align}
i_d^{min} = -\frac{ L_dK }{ L_d^2 + R (n_p \omega_{MN} k_{Fe})^{-1}  }         ,
\end{align}
which is independent of quadrature current $i_q$ if the reluctance torque and the saliency are neglected, what is acceptable for surface-mounted PMSMs \cite{Eisenmodell}. The value is doubled to enable further field-weakening to improve dynamics in high speed, an effect described in the results section. For the quadrature current $i_q$, the largest possible range of values should be available. The resulting linear constraints, shown in Fig. \ref{fig:constraints}, almost completely fill the current region of interest. A representation with affine inequality constraints is thus acceptable.

The approximation of the voltage constraints is a bit more difficult. The $q$-axis should not be restricted, as the induced voltage is aligned to it and is the largest value that will appear. A steady-state analysis of the system equations (2), (3) shows that a rectangular voltage area results
\begin{align}
 R i_{d}^{min} \mbox{-} n_p L_q \omega_{M}^{max} i_{q}^{max}                &\le   u_d    \le   n_p L_q \omega_{M}^{max} i_{q}^{max}  ,\\
\mbox{-}R i_{q}^{max}  \mbox{+}n_pL_d\omega_{M}^{max} i_{d}^{min}  \mbox{-}n_p K \omega_{M}^{max}  &\le   u_q    \le   R i_{q}^{max} \mbox{+}n_p K \omega_{M}^{max} .
\end{align}
This rectangle (dark grey on Fig. \ref{fig:constraints}) is expanded such that the outer circle of the voltage limitation is hit (light grey on Fig. \ref{fig:constraints}). During dynamical transients, the voltage vector points to one of the outer corners, subsequently touching the outer limiting circle. Therefore, a linear approximation of the voltage limits as a rectangle by the presented method, as shown on Fig. \ref{fig:constraints}, does not limit the steady-state operational range and only marginally affects dynamics. A less restrictive method is presented in \cite{onlineMPC}, where a time-varying constraint in form of a hexagon in stator frame is proposed. The limiting circle can be expanded to a hexagon by using overmodulation techniques \cite{Quang}. While this is possible too with the underlying predictive control algorithm, the method in the ($d,q$)-frame is chosen for simplicity and to prevent possible current ripples.

\psfrag{#iq}{$i_q$}
\psfrag{#id}{$i_d$}
\psfrag{#uq}{$u_q$}
\psfrag{#ud}{$u_d$}
\psfrag{#imx}{$I_{max}$}
\psfrag{#umx}{$U_{max}$}
\psfrag{#-il}{$-\frac{I_{max}}{2}$}

\section{Optimal Control Algorithm}

To study real-time applicability of the presented scheme, first, the highest possible amount of optimization parameters is determined. The fastest optimizer with constraints is the widely known linear programming (LP) method. Table I shows some worst-case computational results of LP (simplex method from \cite{LPalg}) as function of the number of free parameters (CPU: $1.4$ GHz industrial PC). More parameters lead to a higher number of iterations which are also more complex; the worst-case number of iterations is the number of parameters plus the number of constraints \cite{Pierre}. As in the underlying application, the constraints are decoupled, however, this worst-case is not to be expected. The maximum runtime is given by the sampling rate minus latency of input/output, therefore at $8$ kHz sampling rate, it must be less than about $110$ $\mu$s. Thus, at best, $12$ parameters can be optimized if a LP method is used. Runtime of the predictive controller is further discussed at the end of the section.

\begin{table}[!htb]
\renewcommand{\arraystretch}{1.2}
\caption{Runtime of a linear program for some worst-case problems on a $1.4$ GHz CPU}
\label{tbl:computer}
\centering
\begin{tabular}{|c|c|c|c|}
\hline
Parameters   &  Constraints & Iterations &  Runtime [$\mu$s] \\
\hline
  20         &   44         &   67       &   769  \\
\hline
  12         &   28         &   34       &   165  \\
\hline
  8          &   20         &   10       &   35  \\
\hline
\end{tabular}
\end{table}


\subsection{Trajectory Generation}

The trajectory generation algorithm presented in \cite{SK10}, a development related to flatness-based methods \cite{Guay}, is chosen. It can optimize a quadratical cost function with linear constraints. As major differences to standard algorithms, it is applying a continuous parameterization instead discretization, and the computationally efficient linear programming solver is used instead of quadratic programming or iterative gradient search. Even though LP is used, it is still quadratic optimization; the unconstrained solution to the quadratic cost is calculated first, then, constraints are included with the LP solver.

\begin{figure}[!htb]
  \centering
  \includegraphics[width=8cm]{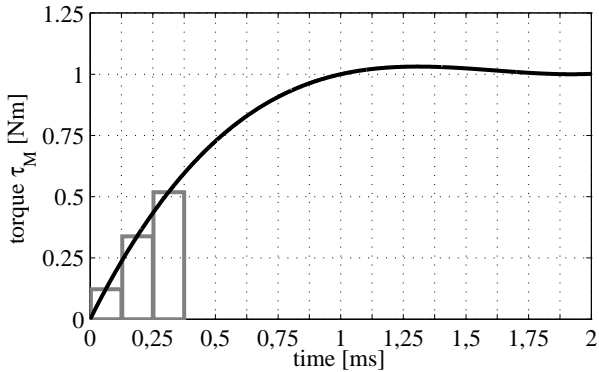}
  \caption{A trajectory described by $3$ free parameters. Black: polynomial-based curve, grey: discrete-time curve. Discrete-time horizon is $0.375$ ms, continuous polynomial-based trajectory is well-conditioned at the desired horizon length $2$ ms.\label{fig:poly}}
\end{figure}

The trajectories for the current are defined as degree $n$ power series with undetermined coefficients $\alpha_{ij}$,
\begin{align}
i_{d}(t) = \sum_{k=0}^{n} \alpha_{dk} \ \frac{t^{k}}{T^k} ,  \;\;\;\;  i_{q}(t) = \sum_{k=0}^{n} \alpha_{qk} \ \frac{t^{k}}{T^k} , \;\;\;\; t\in[0,T]  .  \label{eq:trajectory}
\end{align}
This definition reduces the dimensionality of the generated trajectories rather than their length and is referred to as Ritz parameterization \cite{Pierre}. It is an alternative parameterization to the typical Euler discretization. The first coefficients $\alpha_{d0}$ and $\alpha_{q0}$ are the initial conditions, and the remaining $6$ coefficients are determined by optimization. A high prediction horizon is obtained for a relatively small number of parameters. Due to the analyzed computational limitations, $n=3$ is chosen as polynomial degree. Fig. \ref{fig:poly} illustrates the computational advantage, with $3$ parameters at $8$ kHz sampling rate, using a discrete description, the prediction horizon is $0.375$ ms, but with a degree $3$ polynomial, a well-conditioned setpoint change can be described over the desired prediction horizon of $2$ ms. A more complex trajectory may not be expected for the underlying application.

The corresponding voltages $\bm{u}_{dq}(t)$ are computed by algebraic differentiation of (\ref{eq:trajectory}) and by solving for the model equations (\ref{eq:id}) and (\ref{eq:iq}), this is also called the flatness-based approach \cite{HSR04}. This way, the voltages do not need to be represented by additional parameters, and equality constraints are avoided in the optimization. The variables $\bm{i}_{dq}$, $\ddt \bm{i}_{dq}$ and $\bm{u}_{dq}$ are substituted in the cost functional $J$ (\ref{eq:J}) by the found functionals.

\begin{figure*}[!htb]
  \centering
  \includegraphics[width=16cm]{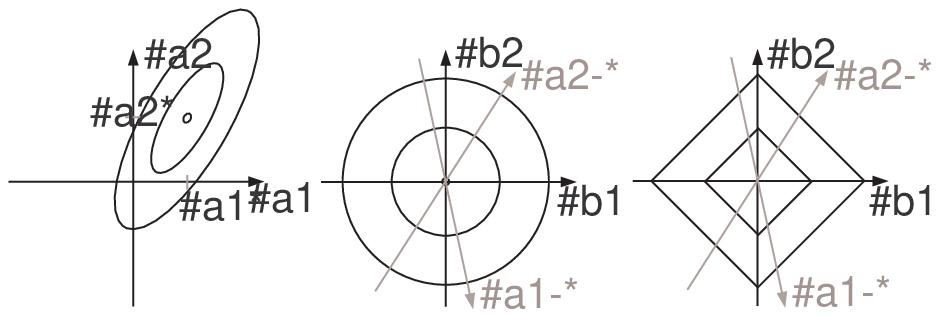}
  \caption{Transformation and linearization of the optimization problem. Lines are equimagnitude contours of the cost function. Left: original problem, middle: transformed least-distance problem, right: transformed and linearized problem.\label{fig:algorithm}}
\end{figure*}

The cost functional $J$ is then a quadratic function of the unknown parameters $\bm{\alpha}$ and of the parameters which were assumed constant, namely the motor parameters, the measured currents, the speed $\omega_M$ and the torque reference $\tau_M^*$. Defining the vector of undetermined coefficients as $\bm{\alpha}=(\alpha_{d1},\alpha_{d2},\alpha_{d3},\alpha_{q1},\alpha_{q2},\alpha_{q3})^T$, it is written as
\begin{align}
J = \bm{\alpha}^T \bm{Q} \bm{\alpha} + \bm{q}^T\bm{\alpha} + q_0 .
\end{align}
Because of the parameterization with a polynomial basis, convexity must be discussed \cite{Guay}, in (\ref{eq:J}) the weight matrix was $\mathbb{R}^{2x2}$ whereas now it is extended to $\bm{Q} \in \mathbb{R}^{6x6}$. Convexity guarantees the existence of a unique global minimum, and convergence of the optimization algorithm in finite time. The proof that convexity is maintained with this transformation is given in \cite{SK10}. Graphically, for a simple example with two parameters, $J$ can be represented as in Fig. \ref{fig:algorithm} (left).

The inequality constraints are also parameterized with the polynomial. Exact parameterization requires linear matrix inequality (LMI) methods, which are however too involved for this application. A simpler way is to sample the trajectories for $\bm{i}_{dq}(t)$ and $\bm{u}_{dq}(t)$ at an interval $\frac{T}{n}$. In \cite{SK10} it is proven that if an additional interlay is added, the transformation guarantees maintenance of the original constraints. For instance, a constraint $i_d(t)\le 0 \forall t\in[0,T]$ is parameterized as  
\begin{align}
i_d(0) &\le 0  , \\
i_d(k\frac{T}{n}) - \Delta i_d(0) &\le 0 \; k = 1..3 , 
\end{align}
where $\Delta = 0.064$ is the respective interlay constant for $n=3$ (see \cite{SK10}). The first constraint on the initial condition does not need to be included, and the remaining $3$ conditions are affine functions of $\bm{\alpha}$, but not of $t$, such that they can directly be included in linear-quadratic optimization in the parameter space $\bm{\alpha}$.


As $J$ is convex, the unconstrained global optimum $\bm{\alpha}_0^*$ is found algebraically by solving first-order necessary conditions,
\begin{align}
\bm{\alpha}_0^* = -\frac{1}{2} \bm{Q}^{-1}\bm{q}.
\end{align}
Then, by an affine coordinate transformation
\begin{align}
\bm{\beta} = \bm{A}(\bm{\alpha}-\bm{\alpha}_0^*) , \label{eq:trafo}
\end{align}
the problem can be reformulated as least-distance problem, i.e. a quadratical cost describing the distance to the unconstrained optimum. The linear transformation includes a shift of the origin of the coordinates, a coordinate rotation as well as a coordinate scaling, and is found with
\begin{align}
\bm{A}^T\bm{A} = \bm{Q} ,
\end{align}
which can be solved with the Cholesky decomposition $\bm{A}^T = \textnormal{cholesky}(\bm{Q}^T)$. As result, the cost functional looks much simpler and reduces to a sum of squares
\begin{align}
J = \bm{\beta}^T \bm{\beta} ,
\end{align}
and can be represented as in Fig. \ref{fig:algorithm} (middle). The unconstrained optimum is thereby $\bm{\beta}_0^*=\bm{0}$. This constrained least-distance problem is already simpler to solve than the original problem. In the next step, the least-distance problem is linearized around the unconstrained optimum, see Fig. \ref{fig:algorithm} (right). The squares in the cost function are replaced by absolute values
\begin{align}
J = \bm{\beta}^T \bm{\beta} = \sum_i \beta_i^2 \approx \sum_i | \beta_i | = J' .
\end{align}
In the LP standard form, furthermore, only positive parameters are possible, therefore the variables are replaced by $\beta_i = \beta_{ip}-\beta_{in}$, with $\beta_{ip}, \beta_{in}\ge 0$. The absolute value can then be replaced by $|\beta_i| = \beta_{ip}+\beta_{in}$. Equivalence is guaranteed by minimizing the (positive) sum, such that at least one variable of each pair $(\beta_{ip},\beta_{in})$ will be zero \cite{Pierre}. The linearization of the cost function inherits a large error in the value of $J$, but the values of the coefficients $\beta$ are not affected that much: the least-distance problem is not so much different in the linear form as it would have been in the quadratical form. Furthermore, see that a difference only appears if a constraint is active, the unconstrained optimum is the same. Suboptimality is evaluated in the results section based on numerical simulations. The linear constraints on $\bm{\alpha}$ are as well transformed directly with (\ref{eq:trafo}). Therefore, after all the transformations, the problem is available in standard form for linear programming, and a simplex solver \cite{LPalg} can be run. The optimal solution $\bm{\beta}^*$ has to be retransformed to find the optimum in the original coordinates
\begin{align}
\bm{\alpha}^* = \bm{\alpha}_0 + \bm{A}^{-1}\bm{\beta}^*   ,
\end{align}
which is used in (\ref{eq:trajectory}) and the motor model to obtain the optimal trajectories for the currents and voltages.



\psfrag{#a1}{$\alpha_1$}
\psfrag{#a2}{$\alpha_2$}
\psfrag{#b1}{$\beta_1$}
\psfrag{#b2}{$\beta_2$}
\psfrag{#a1*}{\textcolor[gray]{0.5}{$\alpha_1^*$} }
\psfrag{#a2*}{\textcolor[gray]{0.5}{$\alpha_2^*$} }
\psfrag{#a1-*}{\textcolor[gray]{0.5}{$\alpha_1$-$\alpha_1^*$} }
\psfrag{#a2-*}{\textcolor[gray]{0.5}{$\alpha_2$-$\alpha_2^*$} }

\subsection{Implementation Aspects}

All computations of the previous section are done using a computer algebra tool (Maplesoft Maple). The matrices for the LP solver are generated using the C code generation toolbox. The real-time software thus consists of a simplex tableau assignment, consisting of the initialization of one matrix and two vectors of floating-point variables, which is automatically generated code, a simplex LP solver from \cite{LPalg}, and some post-processing. As the assignment is based on symbolic calculations, the motor parameters can be changed online. However, the majority of parameters is defined as constant, the compiler can then optimize code size as well as runtime of the initialization stage.

Alternatives to this procedure were evaluated. The direct use of a quadratic program is too slow, for instance, with $6$ parameters, $16$ (voltage) constraints and $20$ iterations it requires $730$ $\mu$s - if additionally current constraints were included, it would take even longer. If the problem is first transformed to least-distance, the number of iterations and the runtime reduce to about $50\%$. 

Also online calculation of the trajectory generation algorithm instead of computer algebra generated code was analyzed. The matrix inversion with $6$ parameters takes $2.5$ $\mu$s and the cholesky decomposition $40$ $\mu$s. Additionally this requires the use of a scientific mathematics library.

The choice to use a computer algebra tool and automatic code generation turns out as the easiest way to implement and as well as computationally most efficient. The automatically generated code is copy-and-pasted to the real-time software, only a small LP solver like \cite{LPalg} has to be added. The chosen LP implementation, however, seems to be a somewhat slow implementation, runtime improvements are possible.

\subsection{Predictive Control}

\begin{figure*}[!ht]
  \centering
  \includegraphics[width=17cm]{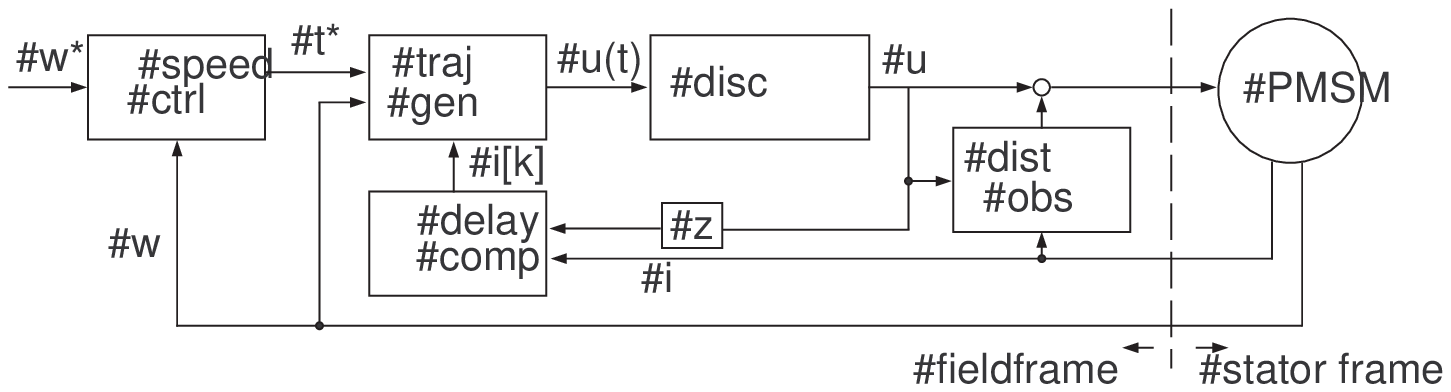}
  \caption{Control structure of the predictive torque controller cascaded by PI speed control.\label{fig:structure}}
\end{figure*}

The control structure is shown in Fig. \ref{fig:structure}. A cascaded control structure is chosen as speed is assumed constant for trajectory generation. As the mechanical plant is generally only roughly known, this overlying speed controller is advantageous. Model-based control can be used for the electrical subsystem of the motor as the parameters are known, but for the mechanical part, any robust feedback controller can be chosen.

The trajectory generation scheme is embedded in a predictive controller and repeated at every sampling step to generate the optimal control input $\bm{u}_{dq}[k]$.

Additionally, for steady-state accuracy, a disturbance observer is necessary, for instance to compensate modeling errors of the induced voltage terms. Even though the induced voltage is modeled, variations in $K$ caused for instance by heating or identification errors cause an offset of the current $i_q$. The approach is to assume the disturbance as constant and calculate it directly based on the MPC model and past control inputs and current measurements \cite{disturbance}. The respective delays have to be accounted in the design. 

\psfrag{#speed}{speed}
\psfrag{#ctrl}{control}
\psfrag{#traj}{trajectory}
\psfrag{#gen}{generation}
\psfrag{#delay}{delay}
\psfrag{#comp}{comp.}
\psfrag{#disc}{discretization}
\psfrag{#PMSM}{PMSM}
\psfrag{#dist}{disturbance}
\psfrag{#obs}{observer}

\psfrag{#fieldframe}{rotor coordinates}
\psfrag{#stator frame}{stator coordinates}

\psfrag{#w}{$\omega_M$}
\psfrag{#w*}{$\omega_M^*$}
\psfrag{#t*}{$\tau_M^*$}
\psfrag{#u(t)}{$\bm{u}_{dq}(t)$}
\psfrag{#u}{$\bm{u}_{dq}[k]$}
\psfrag{#i[k]}{$\bm{i}_{dq}[k|k$-$1]$}
\psfrag{#i}{$\bm{i}_{dq}[k$-$1]$}
\psfrag{#z}{$z^{-1}$}

\subsection{Timing}

\begin{figure*}[!ht]
\centering
\subfigure[Timing at $0$ rpm.]{\includegraphics[width=7.8cm]{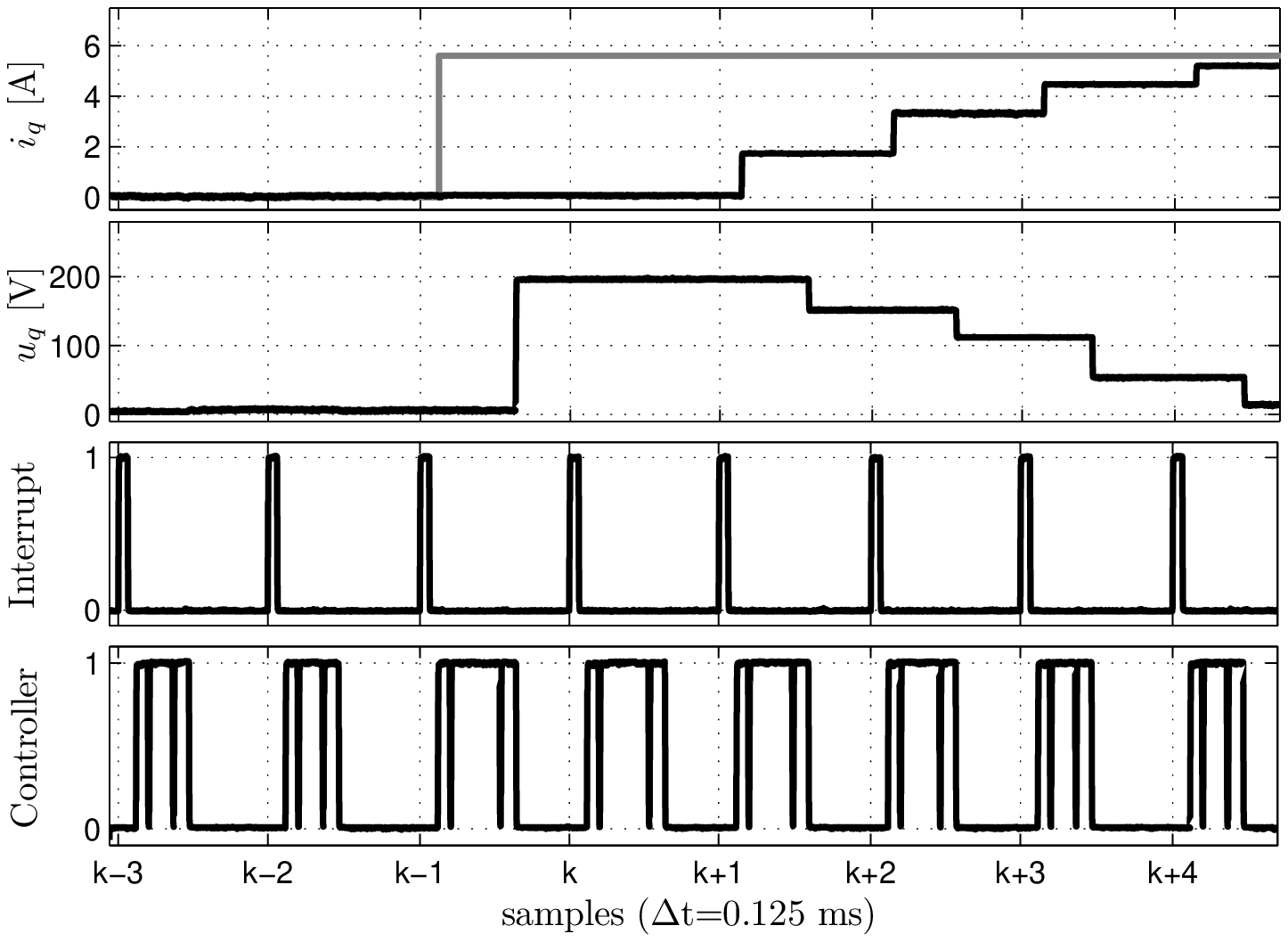}}
\subfigure[Timing at $2000$ rpm.]{\includegraphics[width=7.8cm]{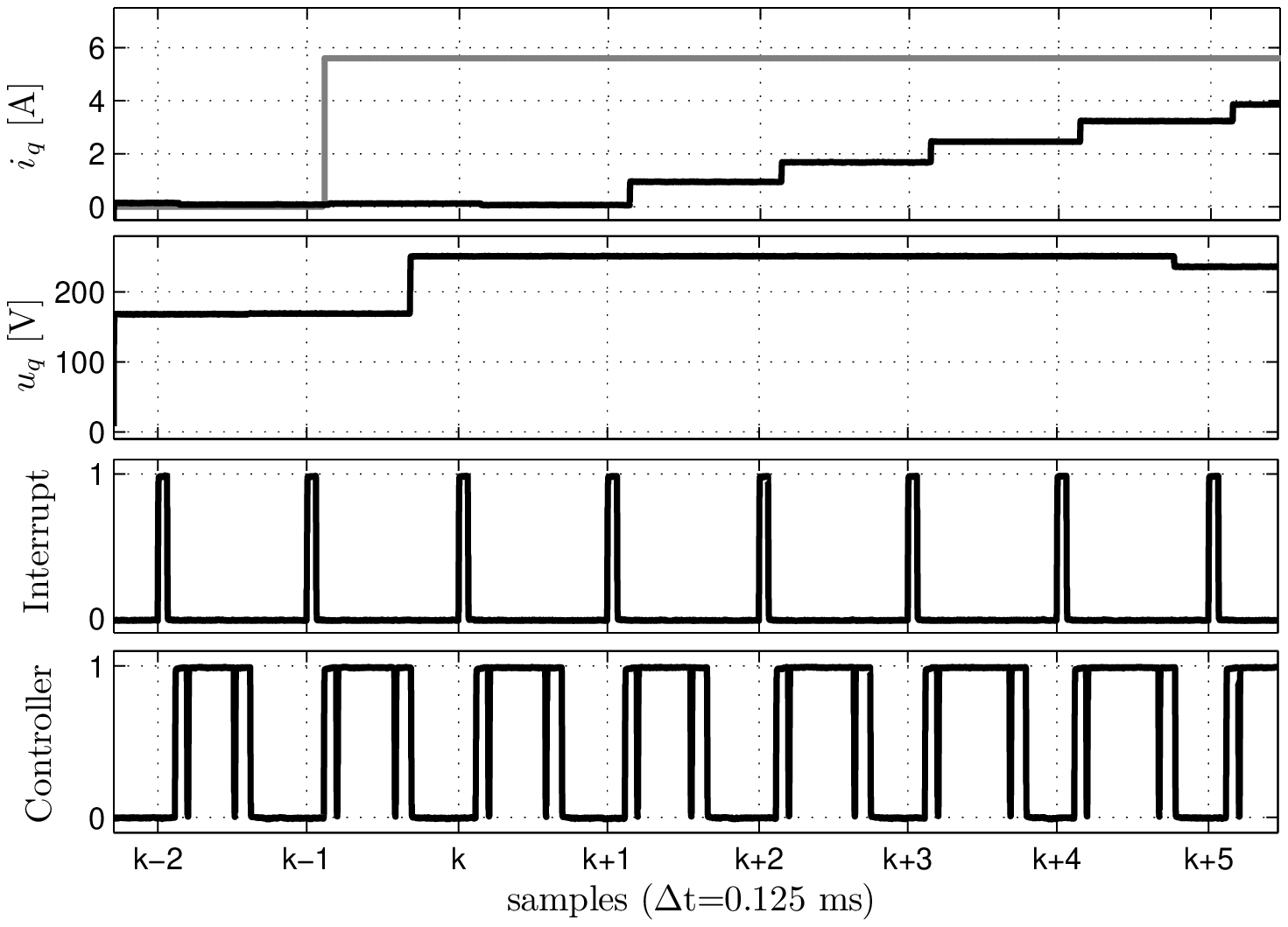}}
\caption{Experimental timing results of the predictive control scheme. 1: quadrature current $i_q$ without delay compensation, 2: quadrature voltage $u_q$, 3: interrupt handling, 4: control law computation (first part: initialization, second part: LP solver, third part: post-processing).\label{fig:timing}}
\end{figure*}

First, measurement and control timing is analyzed. Two timing sequences are shown in Fig. \ref{fig:timing} (a) and (b). The interrupt-based control system triggers an interrupt every $125$ $\mu$s (signal 3). At this instant, the applied voltage command (signal 2) is modulated in the power electronics. At the same instant, the A/D conversion of the current measurements (signal 1) is performed to avoid the impact of current ripples. It is seen that to calculate the command $u_q[k]$, only the current $i_q[k-1]$ is available. Furthermore, once the command $u_q[k]$ is calculated, it needs to wait until the next interrupt. This delay of one step is denoted \textit{computational delay} and is accounted with a delay compensation technique. A prediction of one sampling step with the model, the current $i_q[k-1]$ as well as the previously commanded voltage $u_q[k-1]$ is applied to generate the delay-compensated current $i_q[k|k-1]$ as initial condition for the control law to compute $u_q[k]$, see \cite{Moon}. Furthermore, see that the response to the commanded voltage $u_q[k]$ is $i_q[k+1]$ which is available one interrupt later, this is the \textit{plant delay} and it is naturally included in the predictive controller by recalculating the trajectory at every sampling step. From the predicted trajectory, the applied voltage is $\bm{u}_{dq}[k] = \bm{u}_{dq}(0)$.

From the timing sequences, interesting insight into the computational demands of the algorithm is gained. The first part of the controller signal in Fig. \ref{fig:timing} shows the calculation time for the simplex tableau initialization, it takes about $10\mu$s. Included in these calculations, which is automatically generated code resulting from symbolical calculations, is a calculation of the unconstrained optimum and the linearization of the problem. The second and biggest part of the controller signal is the runtime of the linear program \cite{LPalg}. At the beginning of Fig. \ref{fig:timing} (a), where voltage and current are both zero, it is only about $20\mu$s, but to calculate the voltage step at $2000$ rpm shown in Fig. \ref{fig:timing} (b), more iterations are involved as many constraints are active, and the computation time rises to almost $60\mu$s. The total time of the interrupt handling, latency, the simplex initialization, the LP solver and the post-processing sum up to almost $100\mu$s in the worst case, therefore up to $80\%$ of the available time is used.

\section{Results}

A surface-mounted PMSM with parameters shown in table II is used. It is coupled to a load drive such that an arbitrary speed can be imposed. The algorithm is implemented on a PC-104 based real-time system with a $1.4$ GHz CPU described in \cite{Nael}. The voltage limitation was set to $250$ V because of hardware limitations, accordingly, rated speed is reduced from $3000$ to $2200$ rpm.

\begin{figure*}[!ht]
\centering
\subfigure[Speed steps from $0$ to $1000$ to $2000$ rpm.]{\includegraphics[width=7.8cm]{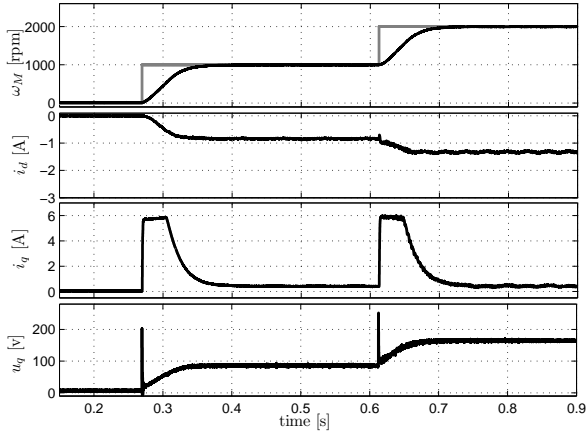}}
\subfigure[Torque step at zero speed.]{\includegraphics[width=7.8cm]{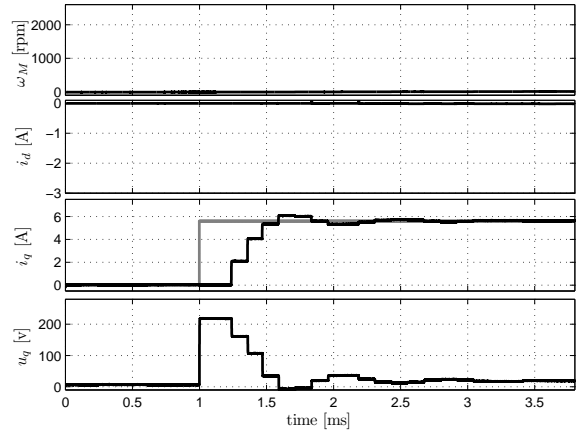}}
\subfigure[Torque step at $2000$ rpm.]{\includegraphics[width=7.8cm]{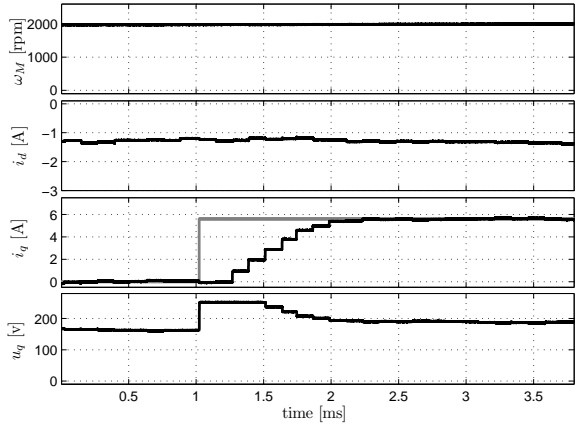}}
\subfigure[Torque step at $2400$ rpm.]{\includegraphics[width=7.8cm]{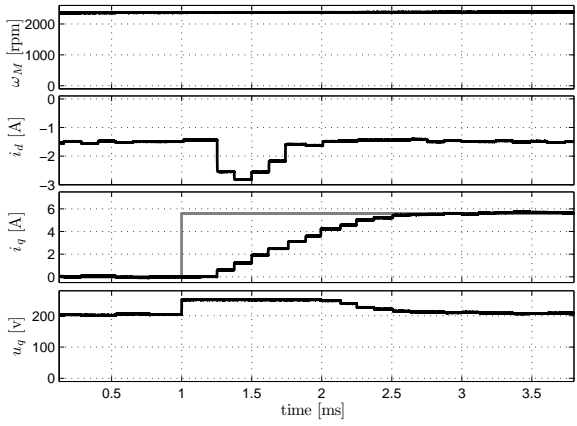}}
\caption{Experimental results of the predictive control scheme. 1: rotor speed $\omega_M$, 2: direct current $i_d$ without compensation, 3: quadrature current $i_q$ without compensation, 4: quadrature voltage $u_q$.\label{fig:results}}
\end{figure*}

Experimental results of the proposed scheme are shown in Fig. \ref{fig:results}. Subfigure (a) shows the response to two subsequent speed reference steps, the load drive is deactivated. The torque is increased rapidly by a high but feasible voltage peak - the PI speed controller can be made very fast such that also a step appears on the reference torque for the MPC controller. The direct current depends on the speed and thereby reduces iron losses which are considerable at high speeds. Losses are decreased by about $4\%$, and the efficiency is improved by about $0.5\%$ at $2000$ rpm. Better results are obtained on motors with higher inductances \cite{PMSMeff}.


The next three subfigures (b), (c) and (d) show fast torque transients at zero, medium and high speed, respectively. The PMSM is in torque control mode while the load drive keeps speed constant at $0$, $2000$ or $2400$ rpm, respectively. The current components are well decoupled, a fast current change on the quadrature axis does not affect the direct axis in (b) and (c). With two separate PI controllers, a short current excursion would be seen on the direct axis during the torque transient. Again, the current on the direct axis $i_d$ is dependent on the speed. Furthermore, the torque change is fast and at the same time smooth, the voltage becomes smoothly small for smaller control errors -- a nice characteristic of quadratic cost functionals, compared to linear cost functions which result in deadbeat behavior and are more sensitive to uncertainties \cite{Moon}. 

On subfigures (b) and (c), the behavior with active voltage constraint is the same as when using standard saturation or anti-windup strategies. On subfigure (d), however, a different behavior is seen, the direct current $i_d$ is reduced to perform field-weakening. This implies that the stator induced voltage is reduced on the quadrature axis, see eq. (\ref{eq:iq}). This mechanism is known from minimum-time current control \cite{Sul}. With field-weakening, the gap between induced and maximum voltage increases, the derivative of the quadrature current $\ddt i_q$ is higher and the torque-generation dynamics are increased, at the cost of higher copper losses on the direct axis. Without additional field-weakening, the reference torque would not be reached after the optimization horizon of $2$ ms, thereby the end-weight of the control error in $J$ oversizes the loss term. Therefore, in this predictive control implementation, field-weakening not only improves efficiency, but also improves dynamics by exploiting cross-coupling between the orthogonal current components to optimally bypass the voltage saturation. 

\begin{figure*}[!ht]
\centering
\subfigure[Torque step at $2000$ rpm.]{\includegraphics[width=7.8cm]{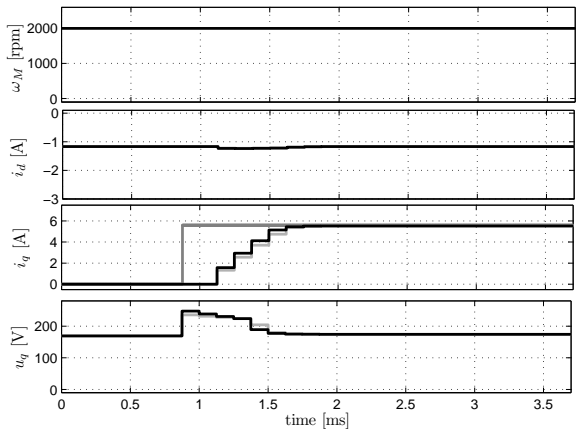}}
\subfigure[Torque step at $2400$ rpm.]{\includegraphics[width=7.8cm]{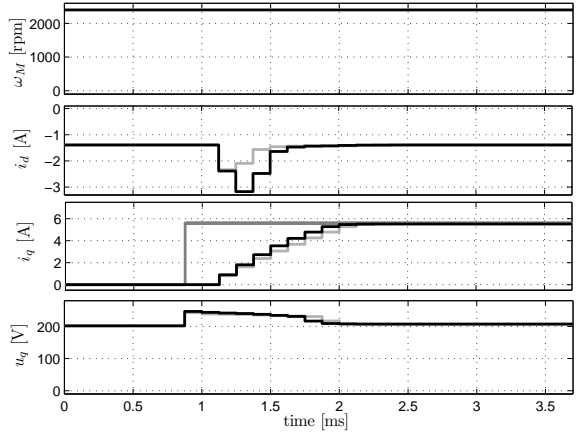}}
\caption{Simulation results of the predictive control scheme, black: QP, light grey: LP. 1: rotor speed $\omega_M$, 2: direct current $i_d$ without compensation, 3: quadrature current $i_q$ without compensation, 4: quadrature voltage $u_q$.\label{fig:simresults}}
\end{figure*}

The situation described above is the one where the highest number of constraints is active, and thus, also where the suboptimality of the proposed trajectory generation method is the highest. Based on the simulation results in Fig. \ref{fig:simresults}, the suboptimality can be analyzed by comparing the results to a QP solver, which is not practicable in the experimental setup. In Fig. \ref{fig:simresults} (a) no real difference can be seen as only few constraints are active. In Fig. \ref{fig:simresults} (b) more constraints are active, as result, it is seen that the dynamical field-weakening is much more intense, but foremost, of longer duration when using QP. The simulations in Fig. \ref{fig:simresults} have a slightly different quantitative behavior as the experiments, caused by parameter uncertainties.

It is also possible to operate the PMSM beyond rated speed with steady field-weakening to bypass the voltage saturation on the quadrature axis, as shown in \cite{inhFW}. It is remarked that the current on the direct axis $i_d$ has no reference, its value is obtained from the optimization of the cost functional. Therefore the method works well and is numerically stable; the optimal value follows inherently.

\section{Conclusion}

Two questions have been central to this work, first, is it possible to implement linear model predictive control (MPC) with online optimization on an electrical drive,  and second, will this advanced controller lead to any merits.

A model predictive control scheme for a PMSM was introduced. Based on suboptimal real-time optimization, the currents and voltages are computed according to a cost functional. The prediction horizon is $2$ ms at a sampling rate of $8$ kHz, and voltage and current constraints are respected. The work can be seen as first implementation of classical model predictive control on an electrical drive, where the aspects of constraint inclusion and high prediction horizon are satisfied while online optimization is used. Implementing online MPC is therefore possible even on fast-sampling systems such as electrical drives.

Additionally to the advantages of general predictive controllers, which are precise accounting for timing of measurement and control as well as respecting current and voltage constraints, the advantages of the long-range constrained predictive MIMO control scheme can be concluded as follows: improved accounting of cross-coupling, fast and smooth dynamical behavior, improved power efficiency by field weakening, and improved dynamics close to voltage saturation by field-weakening.

As negative point, the high demands to computational power must be named.

\section*{Acknowledgments}
This work was supported through the National Research Fund of Luxembourg under grant PhD-08-070. The authors thank Janos Jung for his valuable help at the implementation stage and Sascha Kühl for the fruitful discussions.


\section*{Appendix: Motor Parameters}

\begin{table}[!h]
\renewcommand{\arraystretch}{1.2}
\caption{Nominal Parameters of the Synchronous Motor\label{tbl:SMparameter}}
\centering
\begin{tabular}{|l||c|}
\hline
Manufacturer $\&$ Model  & Merkes MT5 1050   \\
\hline
Rated Power $P_N$         &  $2760$ W  \\ 
\hline
Rated Torque  $\tau_{MN}$ &  $10.5$ Nm  \\
\hline
Rated Current   (peak)      &  $8$ A \\
\hline
Rated Speed $\omega_{MN}$ &  $3000$ rpm \\
\hline
Pole Pairs  $n_p$         &  $3$ \\
\hline
Rated Voltage $U_{N}$   (peak)  &  $560$ V resp. $330$ V\\
\hline
Stator Inductance $L_d$, $L_q$     &  $4.8$, $7.2$ mH \\
\hline
Stator Resistance $R$     &  $0.92$ $\Omega$ \\
\hline
Motor Constant $K$  (peak)   &  $0.334$ Vs \\
\hline
Iron  Loss Constant (Hysteresis) $k_{Fe}$        &  $1.27$ $\frac{\textnormal{A}}{\textnormal{Vs}}$ \\
\hline
\end{tabular}
\end{table}

\end{document}